# МОДЕЛИРОВАНИЕ КЛАСТЕРНЫХ СТРУКТУР В МАТЕРИАЛЕ: СИЛОВЫЕ ПОЛЯ И ДЕСКРИПТОРЫ

А.Н. Герега

Одесская государственная академия строительства и архитектуры,
Одесса, 65029, Украина

Проведено моделирование силовых полей интерьерных границ вещества как суперпозиции полей модифицированных квадратов Серпинского.

Установлено ограничение на применимость коэффициентов поврежденности материала; предложен эффективный информационный дескриптор интерьерных границ.

Введено представление об «эффекте дальнодействия» в облученных кристаллах как о различных проявлениях фазового перехода второго рода. Предложена феноменологическая перколяционная модель явления, приведена интерпретация экспериментальных данных, обсуждается возможный механизм эффекта.

*Ключевые слова:* интерьерные границы, силовое поле, квадрат Серпинского, перколяция, эффект дальнодействия, фрактальный кластер, структурный фазовый переход, информационный дескриптор

# MODELING OF THE CLUSTER STRUCTURE OF MATERIALS: FORCE FIELDS AND DESCRIPTORS

A.N. Herega

Odessa State Academy of Civil Engineering and Architecture, Odessa, 65029, Ukraine

The simulation of the force fields of interior borders matter as superposition of the modified Sierpinski carpet was realized. Studies were shows restriction on the applicability of the coefficients of the material damage, and were offered the effective information descriptor interior boundaries. Ion implantation "long-range effect" is interpreted as phenomena of structural phase transition. Given the interpretation of experimental data, was discussed the possible mechanism of the effect.

*Keywords:* inner boundaries, force field, Sierpinski carpet, percolation, long-range effect, fractal cluster, structural phase transition, information descriptor

## 1. Введение

Материал и внутренние границы – взаимообусловленные и совместно развивающиеся кластерные системы. Перераспределяя деформации в материале, интерьерные границы эволюционируют, изменяя характерные размеры и «осваивая» новые масштабы, тем самым, модифицируя материал. Хорошо известны сложности, возникающие при расчете силовых полей сетей трещин и внутренних границ. В статье предложена модель произвольных сетей интерьерных границ в виде суперпозиции модифицированных предфракталов Серпинского, позволившая получить итерационные алгоритмы для расчета силовых полей.

Регулярно используемые для описания состояния поверхности материала коэффициенты поврежденности являются важными интегральными характеристиками дефектности. В работе определены пределы применимости коэффициентов поврежденности, и рассмотрена возможность описания интерьерных границ с помощью предлагаемого информационного дескриптора – относительной степени упорядоченности внутренних границ.

Ряд физических эффектов в веществе, как известно, объясняется существованием связ-ных (или квазисвязных) областей перколяционного типа [1-3]. Размерность, лакунарность, степень анизотропии и разветвленности и другие параметры таких кластерных систем существенно влияют на кинетические коэффициенты, приводят к аномальной упругости материала и другим эффектам. В статье предложена феноменологическая перколяционная модель так называемого «эффекта дальнодействия», возникающего в кристаллических материалах при ионной имплантации. В модели рассматривается возможный механизм явления, который позволяет интерпретировать ряд хорошо установленных фактов: немонотонную зависимость микротвердости образца от дозы облучения, пороговый характер эффекта, независимость от сорта имплантируемых ионов и вида материала, и другие.

## 2. Модель силового поля мультимасштабной сети интерьерных границ

Многочисленные вариации сетей интерьерных границ поддаются спонтанной визуальной классификации: в замысловатых и неповторимых рисунках их структуры легко угадываются закономерности [4]. Предлагаемая модель силового поля базируется на статистическом подобии сетей границ и предфракталов Серпинского и их модификаций. Она строится в предположении, что любая сеть внутренних границ может быть получена с наперед заданной точностью наложением необходимого количества модификаций регулярных предфракталов произвольных поколений, по аналогии с тем, как любая функция может быть разложена в ряд Фурье.

Определим аналитически силовое поле, создаваемое полимасштабной сетью внутренних границ модификацией квадрата Серпинского, не имеющей осей симметрии (рис. 1).

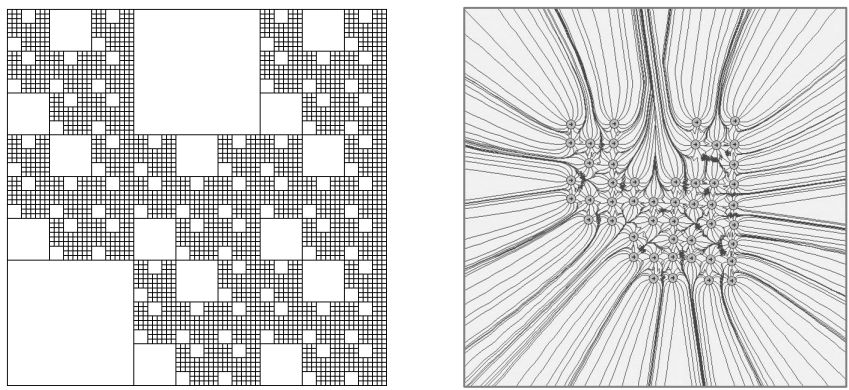

Рис. 1. Несимметричная модификация ковра Серпинского
и ее силовое поле.

Рассмотрим «проволочную» модель несимметричной модификации ковра Серпинского. Пусть исходная квадратная рамка разделена четырьмя «проволоками» на девять равных квадратов. Процедура многократно повторяется на каждой из $7^m$ получаемых на очередном шаге рамок. Пусть также на каждой образующей рамок любого «поколения» с линейной плотностью $\lambda$ содержатся точечные источники, создающие поля напряженностью $E \sim 1/r^2$.

Пусть ковёр Серпинского с длиной стороны образующего квадрата, равной $2H$, расположен так, что его центр совпадает с началом координат, а стороны параллельны осям. Составляющие вектора напряжённости, создаваемой отрезком, определятся соотношениями

$$\begin{cases} E_x = \lambda\, (sin\alpha_2 - sin\alpha_1)\,/\,r, \\ E_y = \lambda\, (cos\alpha_2 - cos\alpha_1)\,/\,r, \end{cases}$$

где $\alpha_i$ – угол между перпендикуляром длиной r, опущенным из точки, в которой определяется напряжённость, на отрезок или его продолжение, и соответствующим направлением на концевые точки отрезка.

Обозначим

$$A(u, v) = (u^2 + v^2)^{-1/2}, \ B(u, v) = v/[u \ (u^2 + v^2)^{1/2}],$$

$$\xi(n, p) = -\xi + (-1)^n p, \ \eta(n, p) = -\eta + (-1)^n p,$$

тогда составляющие вектора напряжённости, создаваемой ковром $m$-го поколения в произвольных точках, не лежащих на прямых, которые содержат отрезки сети, можно вычислить по рекуррентным соотношениям:

$$X_1(\xi;\eta) = \sum_{i=1}^{2}\sum_{j=1}^{2}\sum_{l=1}^{2}(-1)^i[A(\xi(i,h_1);\eta(j,(2l-1)h_1/3)) -$$

$$-B(\xi(j,(2l-1)h_1/3;\eta(i,h_1))]; \quad Y_1(\xi;\eta) = X_1(\eta;\xi);$$

$$X_n(\xi;\eta) = X_{n-1}(\xi+2h_{n-1};\eta-2h_{n-1}) + \sum_{i=1}^{2}\{X_{n-1}(\xi;\eta(i,h_{n-1})-h_{n-1}) +$$

$$+(-1)^i[A(\xi-h_n;\eta(i,2h_{n-1})-h_{n-1}) - B(\xi(i,2h_{n-1})-h_{n-1};\eta-h_n)] +$$

$$+\sum_{j=1}^{2}[X_{n-1}(\xi(i,2h_{n-1});\eta(j,h_{n-1})+h_{n-1}) + (-1)^i\{A(\xi(i,h_{n-1});\eta(j,h_{n-1})+2h_{n-1}) -$$

$$-A(\xi(i,2h_{n-1})+h_{n-1};\eta(i,h_{n-1})+((-1)^j-1)h_{n-1}) +$$

$$+B(\xi(i,h_{n-1})+((-1)^j-1)h_{n-1};\eta(i,h_{n-1}))\}]\};$$

$$Y_n(\xi;\eta) = Y_{n-1}(\xi+2h_{n-1};\eta-2h_{n-1}) + \sum_{i=1}^{2}\{Y_{n-1}(\xi;\eta(i,h_{n-1})-h_{n-1}) +$$

$$+(-1)^i[A(\xi(i,2h_{n-1})-h_{n-1};\eta-h_n) - B(\eta(i,2h_{n-1})-h_{n-1};\xi-h_n)] +$$

$$+\sum_{j=1}^{2}[Y_{n-1}(\xi(i,2h_{n-1});\eta(j,h_{n-1})+h_{n-1}) + (-1)^{i+1}\{B(\eta(j,h_{n-1})+2h_{n-1};\xi(i,h_{n-1})) -$$

$$-B(\eta(i,h_{n-1})+((-1)^j-1)h_{n-1};\xi(i,2h_{n-1})+h_{n-1}) +$$

$$+A(\xi(i,h_{n-1})+((-1)^j-1)h_{n-1};\eta(i,h_{n-1}))\}]\}; \ n = 2,3...m;$$

$$E_x = E_x(x;y) = X_m(-x;-y), \ E_y = E_y(x;y) = Y_m(-x;-y); \quad h_n = H \cdot 3^{n-m}.$$

Силовые поля, создаваемые полимасштабной сетью внутренних границ квадрата Серпинского и его разновидностью с одной осью симметрии второго порядка, рассмотрены в [5, 6].

### 3. Интерьерные границы и их дескрипторы

Отличающиеся поразительным разнообразием сети интерьерных границ – хороший индикатор структуры и состояния материала. Для их описания можно использовать по-разному определяемые коэффициенты повреждённости, которые дают усреднённую оценку состояния поверхности материала. Как правило, это позволяют судить о дефектности материала, но есть и исключения. Например, коэффициент пологости, определяемый как отношение расстояния между крайними точками внутренней границы к ее длине $K = L_{\text{рас.}}/L_{\text{тр.}}$, не всегда решает задачу: оказывается, что границы, существенно различающиеся визуально, могут иметь равные $K$. Этот эффект объясняется развитой фрактальностью одной из интерьерных границ, именно она нивелирует дескрипторную способность коэффициента (рис. 2).

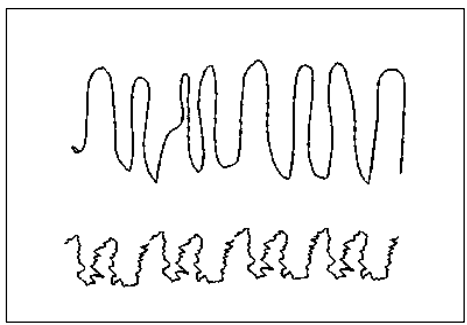

Рис. 2. Эскиз к пояснению потери коэффициентом пологости
трещины дескрипторной способности.

Значительно более эффективный дескриптор основан на определении относительной степени упорядоченности двух интерьерных границ по размерам составляющих их участков. Он базируется на введенном в [7] понятии относительной степени упорядоченности изображений, идеологически связанном с S-теоремой Ю.Л. Климонтовича для открытых систем [8].

Для определения относительной характеристики необходимо обеспечить корректность сравнения. В случае интерьерных границ это можно сделать по аналогии с теоремой Гиббса о сравнении энтропии равновесного и произвольного состояний при условии неизменности среднего значения энергии [9, 10]. В нашем случае это значит, что должно быть обеспечено равенство значений средних длин участков, т.е. одинаковое количество участков в сравниваемых границах должны иметь равную суммарную длину. Для выполнения этого условия по значениям длин последовательно расположенных участков восстанавливаются функции плотности распределения участков по длине $f_1(i)$ и $f_2(i)$, затем по формуле

$$\tilde{f}_2(i) = f_2(i)\left[\sum_i f_1(i) \Big/ \sum_i f_2(i)\right]$$

проводится перенормировка одной из них. В [7] показано, что функционал Ляпунова

$$S_1 - \tilde{S}_2 = -\sum_i \left[f_1(i)\ln f_1(i) - \tilde{f}_2(i)\ln \tilde{f}_2(i)\right] = -\sum_i f_1(i)\ln\left(f_1(i)/\tilde{f}_2(i)\right)$$

для элементов двух равновеликих последовательностей данных есть мера относительной степени упорядоченности, где обозначено: $S, \tilde{S}, f, \tilde{f}$ – энтропии и плотности функций распределения длин участков (реальные и модифицированные соответственно). (Аналогичный результат известен в теории информации как расстояние Кульбака-Лейблера [11], и является мерой различия двух вероятностных распределений).

Возможные варианты оценки относительной степени упорядоченности связаны со сравнением по отношению или разности длин соседних участков границ, по абсолютной величине этой разности и другим. Наиболее чувствительной, по нашим данным, является оценка, сделанная по значениям длин.

Таким образом, величина относительной степени упорядоченности – параметр чрезвычайно чувствительный к локальной структуре материала – обнаруживает себя как эффективный дескриптор внутренних границ.

## 4. К теории «эффекта дальнодействия» при ионной имплантации

Наблюдаемая в кристаллах при ионной имплантации и облучении светом совокупность явлений, за которой закрепилось название «эффект дальнодействия», исследуется в течение, примерно, сорока лет, но, как отмечают авторы [12-14], ни убедительного объяснения в рамках классических моделей радиационной физики твердого тела, ни сколько-нибудь законченной теории, ни единства взглядов на механизмы, лежащие в основе этих явлений, сегодня нет.

Суть эффекта – в изменении структуры материала на расстояниях от облучаемой поверхности, на несколько порядков превышающих длину проективных пробегов ионов (или глубину проникновения энергии облучения). Обычно это проявляется в увеличении плотности дефектов упаковки в объеме кристалла по сравнению с областью пробега ионов [13], возникновении в облученном кристалле аномальной диффузии [15, 16], в немонотонном изменении (по мере набора дозы облучения) микротвердости образца на расстояниях до нескольких сотен микрометров от дефектного слоя [14, 17], и некоторых других.

**4.1. Экспериментальные данные**

Обзор даже основных экспериментальных результатов исследования дефектных структур в имплантированных или облученных кристаллах потребовал бы рассмотрения многих десятков работ [12, 18]. Ниже приведены три эксперимента, характерные для работ по изучению эффекта дальнодействия.

1. В [19] описаны эксперименты по облучению галогенной лампой мощностью 300 Вт фольги из пермаллоя толщиной 20 мкм. В работе обнаружено изменение микротвердости на обеих сторонах фольги. Важно, что интервал доз облучения ограничен сверху и снизу: вне этого интервала микротвердость не изменялась.

Объяснение эффекта авторы [19] строят на аналогии между ионным и световым облучением: «наиболее вероятным является механизм возбуждения ионами упругих (деформационных) волн и их взаимодействие с исходными структурными несовершенствами». Нужно согласиться с авторами: «допущение, что генерация деформационных волн происходит при воздействии света от лампы накаливания, является наиболее трудным пунктом модели [19]».

2. В [17, 20] наблюдалось изменение микротвердости поликристаллических металлов и сплавов толщиной несколько сотен микрометров при дозах облучения ионами $10^{13}$ - $10^{16}$ см$^{-2}$. В работе показано, что величина микротвердости является немонотонной функцией дозы с резко выраженными максимумами, и что явление носит пороговый (по энергии облучения) характер и практически не зависит от сорта ионов, вида металла, толщины фольги, плотности ионного тока и возникает с обеих сторон фольги.

Авторы [21] полагают «ответственными за эффект дальнодействия гиперзвуковые волны». Они считают, что «ион, сталкиваясь с поверхностью, производит своего рода микровзрыв, порождая высокочастотную акустическую волну», и хотя «как правило, волны быстро затухают и сами по себе не могут достичь обратной стороны пластины, но, встречая на своем пути протяженные дефекты, они вызывают их перестройку, которая сопровождается испусканием вторичных волн. Возникает своего рода цепной процесс, который, в конечном счете, способен охватить всю толщу пластины и привести к изменению ее свойств».

3. Недавняя работа [13] существенно дополнила экспериментальные данные об эффекте: при металлографическом исследовании образцов кремния, облученных альфа-частицами с энергией 27.2 МэВ при интегральной плотности потока Ф = $10^{17}$ см$^{-2}$ и проекционной глубине проникновения $R_p$ = 360 мкм, авторы обнаружили не только пять слоев дефектов, расположенных ниже слоя внедрения (380, 423, 627, 720, 764 мкм), но и три – на меньшей глубине (132, 242 и 341 мкм).

Этот результат, устраняющий «односторонность» экспериментальных данных, явился существенным компонентом в череде фактов, позволивших сформулировать представление об эффекте дальнодействия как о различных проявлениях фазового перехода второго рода.

**4.2. Перколяционная модель эффекта дальнодействия**

Предлагаемая перколяционная модель позволяет интерпретировать эффект дальнодействия как результат критического поведения аморфизированного (дефектного) слоя.

Положения модели:

– дефектный слой, расположенный на глубине максимально вероятного проективного пробега ионов, представляет собой квазиплоское несплошное «облако» аморфизированных областей различных размеров;

– при критической дозе облучения, часть областей объединяется, и в кристалле возникает перколяционный кластер аморфизированного слоя [1, 3];

– поле механических напряжений, создаваемое таким слоем, – дальнодействующее, убывающее с расстоянием пропорционально $r^{-b}$, при b, примерно, равном единице.

Разработанная компьютерная программа, реализует континуальную перколяционную модель образования аморфизированного слоя.

В модели координаты центров модифицированных областей, из которых формируется перколяционный кластер, определяются методом Монте-Карло; величину областей можно варьировать, что соответствует изменению энергии имплантируемых ионов. Области считаются соединенными в случае контакта; если они перекрываются, степень разупорядочения возрастает. Предусмотрены четыре степени разупорядочения, а также режимы, «включающие» любое сочетание типов модифицированных областей, т. е. имеется возможность проследить образование перколяционного кластера из всевозможных сочетаний малых кластеров (рис. 3).

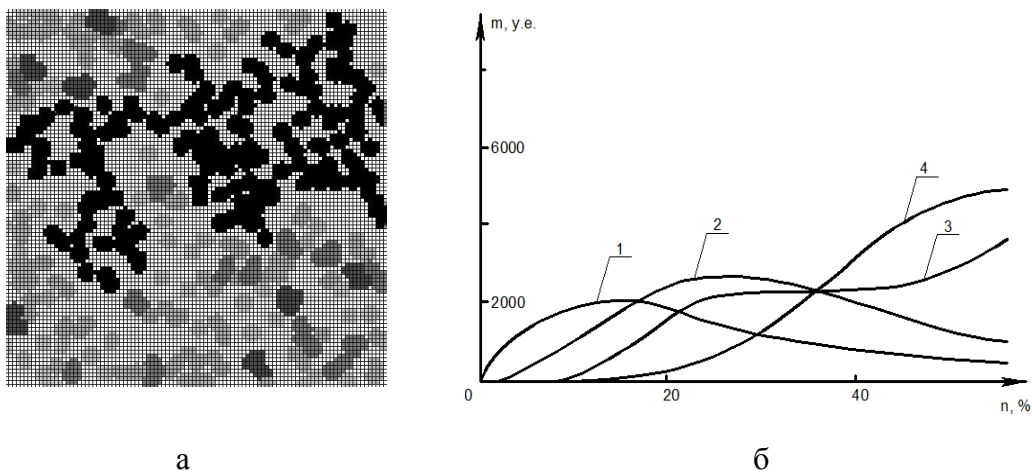

Рис. 3. Модельные области различной степени разупорядочения: а – одна из возможных реализаций; б – кинетика разупорядочения областей дефектного слоя по мере роста его аморфизации. Числа на полочках – степень аморфизации.

В модельных экспериментах определены значения перколяционного и кластерного порогов, радиуса гирации, корреляционной длины, радиус-вектора центра масс кластера, степени анизотропии и первые пять членов спектра размерностей Реньи для перколяционных кластеров, построенных по различным алгоритмам, а также реализована возможность наблюдения динамики процесса построения кластерной системы.

Обсудим кинетику изменения напряжений и возможный механизм явления.

При малых дозах облучения дефектные образования можно рассматривать как отдельные скопления, которые создают механические напряжения, убывающие как $\sigma \sim r^{-3}$ [22]. По мере роста дозы в материале возникает разупорядоченная область перколяционного типа, которая превращается в бесконечный кластер «только при условии, что его плотность превышает некоторую критическую величину [23]». При этом скачкообразно изменяется закон падения напряжений: оно уменьшается заметно медленней, и может быть аппроксимировано законом $\sigma \sim r^{-1}$. При дальнейшем увеличении дозы, перколяционный кластер аморфизированного слоя превращается в «сплошную стенку» с экспоненциально спадающим полем напряжений [22]. Естественно, что столь резкие изменения характера механических полей в материале, существенно видоизменяют равновесное распределение дефектов, и могут приводить к образованию областей с повышенным содержанием дефектов вдали от области вероятнейшего пробега ионов.

Постулирование в модели дальнодействующего характера поля напряжений дает возможность оценить значение фрактальной размерности перколяционного кластера в дефектном слое. Положим, что закон убывания величины механических напряжений с расстоянием имеет вид $\sigma \sim r^{-b}$, где $b = 3 - 2D$, $D$ – размерность дефектов. Тогда при $D = 0$ и $D = 1$, получаются известные зависимости, соответственно, для точечных дефектов ($\sim r^{-3}$) и линейных дислокаций ($\sim r^{-1}$) [22], а величина фрактальной размерности перколяционного кластера в дефектном слое, при которой порождаемое им поле напряжений является дальнодействующим, окажется в пределах $1 < D < 1.5$.

Модель позволяет также интерпретировать немонотонность в зависимости свойств образца от дозы облучения: по мере набора дозы области одной степени разупорядоченности теряют связность, и заменяется перколяционным кластером более разупорядоченных областей. В процессе облучения это происходит неоднократно, и приводит к колебанию величины механических напряжений, к выраженной немонотонности свойств облучаемого образца, в частности, микротвердости, а также позволяет объяснить наличие «магических» доз, при которых наблюдаются экстремумы свойств.

## 5. Заключение

Соотношения для расчета силовых полей могут быть применены ко всему диапазону размеров сетей, в котором проявляется самоподобие. Границы мезоскопической асимптотики могут быть определены по методике, предложенной, например, в [1,24].

Не видно ограничений в возможности обобщения модели на интерьерные границы, расположенные в объеме тела: элементы квадрата Серпинского в «проволочной» модели могут быть заменены на ребра кубов губки Менгера и ее разновидностей.

Возможности предложенного информационного дескриптора определяются масштабом разбиения исследуемого объекта на части, и могут быть повышены до любого разумного предела. (Интересно, что и дифференциальные операторы, примененные к различным параметрам силовых полей интерьерных границ, могут рассматриваться как достаточно чувствительные дескрипторы). По аналогии с рассмотренным может быть построено семейство однотипных дескрипторов по различным характерным параметрам сетей. По другой аналогии, сравнивая такое описание с мультифрактальным, можно представить сеть границ как мультидескрипторный объект, для описания которого нужен бесконечный ряд дескрипторов, аналогичный спектру обобщенных размерностей Реньи.

Помимо описания начальной стадии аморфизации кристаллических тел перколяционная модель эффекта дальнодействия позволяет моделировать кинетику накопления промежуточных веществ и конечных продуктов в последовательности химических реакций, процессы полимеризации и другие.